\documentclass[aps,prb,twocolumn,floatfix,showpacs]{revtex4}

\usepackage[dvips]{graphicx}

\begin{document}

\title{Electron transport in semiconducting carbon nanotubes with 
hetero-metallic contacts}



\author{Yongqiang Xue$^{1,*}$ and Mark A. Ratner$^{2}$}
\affiliation{$^{1}$College of Nanoscale Science and Engineering, University at 
Albany, State University of New York, Albany, New York 12203, USA \\
$^{2}$Department of Chemistry and Materials Research Center, 
Northwestern University, Evanston, Illinois 60208, USA}
\date{\today}

\begin{abstract}
We present an atomistic self-consistent study of the electronic and 
transport properties of semiconducting carbon nanotube  
in contact with metal electrodes of different work functions, which shows 
simultaneous electron and hole doping inside the nanotube junction 
through contact-induced charge transfer. We find that the band lineup 
in the nanotube bulk region is determined by the effective work function 
difference between the nanotube channel and source/drain electrodes, 
while electron transmission through the SWNT junction is affected by 
the local band structure modulation at the two metal-nanotube 
interfaces, leading to an effective decoupling of interface and bulk effects 
in electron transport through nanotube junction devices.      
\end{abstract}


\maketitle


Devices based on single-wall carbon nanotubes 
(SWNTs)~\cite{Dekker,DeMc} have been progressing 
in a fast pace, e.g., the performance of carbon nanotube 
field-effect transistors (NTFET) is approaching 
that of the state-of-the-art silicon Metal-Oxide-Semiconductor 
field-effect transistors (MOSFET).~\cite{AvFET,DaiFET,McFET} But a 
general consensus on the physical mechanisms and theoretical 
models remains to appear. A point of continuing 
controversy in NTFET has been the effect of Schottky barriers at 
the metal-SWNT interface.~\cite{Barrier,AvFET1} Since 
SWNTs are atomic-scale nanostructures in both the axial and  
the circumferential dimensions, any barrier that may form at the 
interface has a finite thickness and a finite 
width.~\cite{AvFET,XueNT,AvFET2} In general 
a microscopic treatment of both the source/drain and gate field 
modulation effect will be needed to account for faithfully the atomistic 
nature of the electronic processes in NTFET.     

Since the characteristics of the NTFETs depend sensitively on the gate 
geometry,~\cite{AvFET} a thorough understanding of the Schottky barrier 
effect in the simpler two-terminal metal-SWNT-metal junction devices is 
essential in elucidating the switching effect caused by applying a finite 
gate voltage.~\cite{XueNT} As a basic device building block, the 
two-terminal device is also of interests for applications in electromechanical and 
electrochemical sensors, where the conduction properties of the SWNT 
junctions are modulated by mechanical 
strain~\cite{NTMe} or molecular adsorption respectively.~\cite{NTCh} 
Previous works have considered symmetric SWNT junctions with different 
contact geometries.~\cite{XueNT} 
Here we consider SWNT in contact with metallic 
electrodes of different work functions. Such hetero-metallic junctions 
are of interests since: (1) The electrode work function 
difference leads to a contact potential and finite electric field 
(built-in field) across the junction at equilibrium. A self-consistent 
analysis of the hetero-metallic junction can shed light on 
the screening of the applied field by the SWNT channel and 
the corresponding band bending effect even at zero bias; (2) For 
SWNTs not intentionally doped, electron and hole doping can be 
induced simultaneously inside the channel by contacting with high 
and low work function metals; (3) Since the metallurgy of the 
metal-SWNT contact is different at the two interfaces, the 
asymmetric device structure may facilitate separating interface 
effect on electron transport from the intrinsic property of 
the SWNT channel. 
 
\begin{figure}
\includegraphics[height=4.0in,width=5.0in]{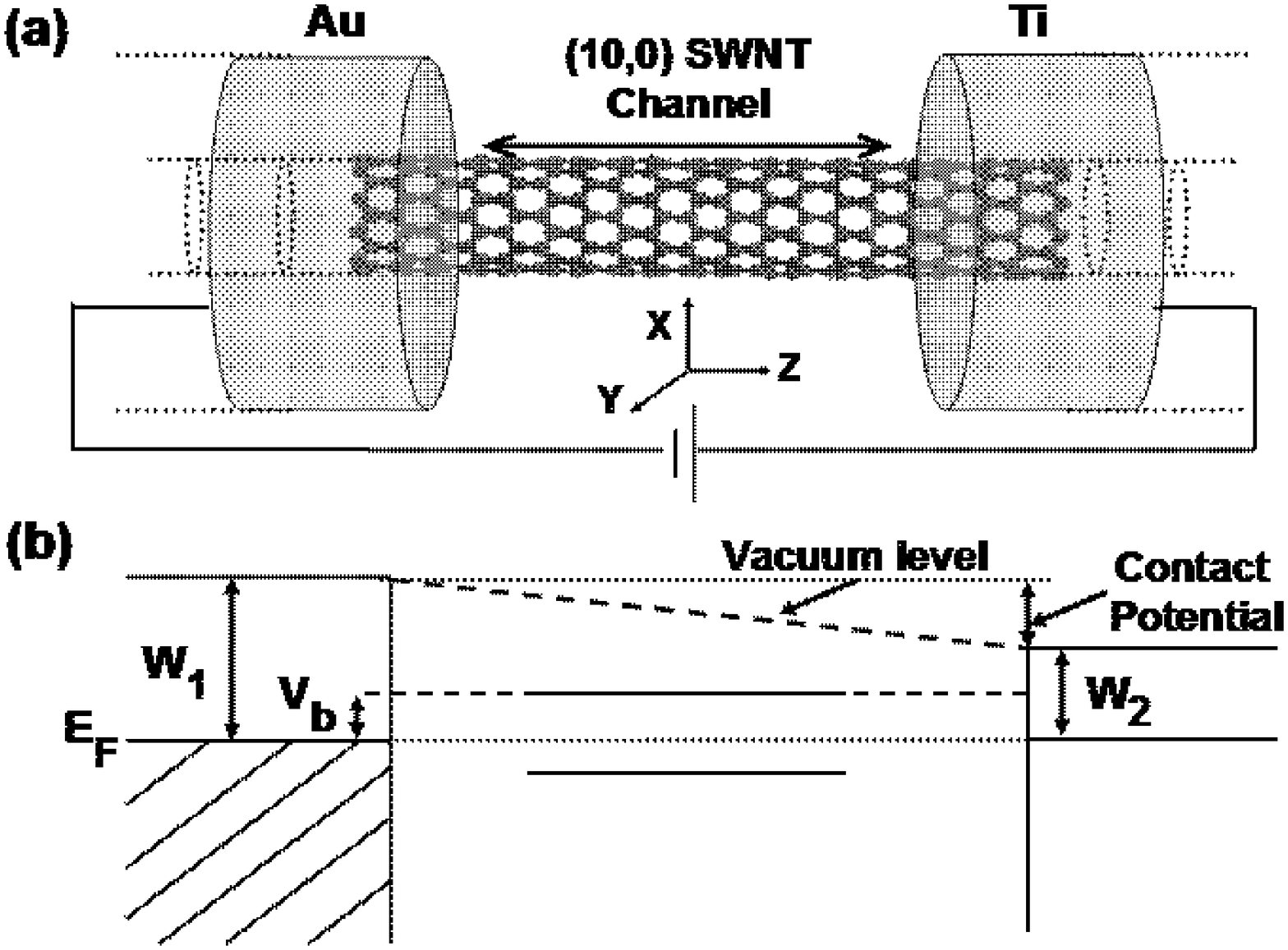}
\caption{\label{xueFig1} (Color online) (a) Schematic illustration of the 
Au-SWNT-Ti junction. The ends of the long SWNT wire are 
surrounded entirely by the semi-infinite electrodes, with only a finite 
segment being sandwiched between the electrodes (defined as the 
channel). Also shown is the coordinate system of the nanotube junction. 
(b) Schematic illustration of the band diagram in the Au-SWNT-Ti junction. 
The band alignment in the middle of the SWNT junction is determined by 
the average of the metal work functions. 
$W_{1(2)},E_{F}$ denote the work functions and Fermi-level of the 
bi-metallic junction. }
\end{figure}

The hetero-metallic SWNT junction is shown schematically in 
Fig.\ \ref{xueFig1}, where the ends of an infinitely long SWNT wire 
are buried inside two semi-infinite metallic electrodes with different 
work functions. The embedded contact scheme is favorable for the 
formation of low-resistance contact. For simplicity, we assume the 
embedded parts of the SWNT are surrounded entirely by the metals 
with overall cylindrical symmetry around the SWNT axis.~\cite{Note1} 
For comparison with previous work on symmetric SWNT junctions, we 
investigate $(10,0)$ SWNTs (with work function of $4.5$ 
eV~\cite{Dekker}) in contact with gold (Au) and titanium (Ti) 
electrodes (with work functions of $5.1$ and $4.33$ eV respectively 
for polycrystalline materials~\cite{CRC}). Choosing the electrostatic 
potential energy in the middle of the junction and far away from the 
cylindrical surface of the SWNT as the energy reference, 
the Fermi-level of the Au-SWNT-Ti junction is the negative of the 
average metal work functions $E_{F}=-4.715$ eV. The SWNT channel 
length investigated ranges from $L=2.0,4.1,8.4,12.6,16.9$ nm 
to $21.2$ nm, corresponding to number of unit cells of $5,10,20,30,40$ 
and $50$ respectively. We calculate the transport characteristics within 
the coherent transport regime, as appropriate for such short 
nanotubes.~\cite{Phonon} 

Using a Green's function based self-consistent tight-binding (SCTB) 
theory, we analyze the Schottky barrier effect by examining the 
electrostatics, the band lineup and the transport characteristics of 
the hetero-metallic SWNT junction as a function of the SWNT channel 
length. The SCTB model is essentially the semi-empirical implementation 
of the self-consistent Matrix Green's function method for \emph{ab initio} 
modeling of molecular-scale devices,~\cite{XueMol} which takes fully into 
account the three-dimensional electrostatics and the atomic-scale 
electronic structure of the SWNT junctions and has been described in 
detail elsewhere.~\cite{XueNT,Note2} The SCTB 
model starts with the semi-empirical Hamiltonian $H_{0}$ of 
the bare $(10,0)$ SWNT 
wire using the Extended Huckel Theory (EHT) with non-orthogonal 
($sp$) basis sets $\phi_{m}(\vec r)$.~\cite{Hoffmann88} 
We describe the interaction between the SWNT channel and the 
rest of the junction using matrix self-energy operators and calculate 
the density matrix $\rho_{ij}$ and therefore the electron density of 
the equilibrium SWNT junction from 
\begin{eqnarray}
\label{GE}
G^{R} 
&=& \{ (E+i0^{+})S-H-\Sigma_{L}(E)-\Sigma_{L;NT}(E)
-\Sigma_{R}(E)-\Sigma_{R;NT}(E)\}^{-1}, \\
\rho &=& \int \frac{dE}{2\pi }Imag[G^{R}](E)f(E-E_{F}).
\end{eqnarray}
Here $S$ is overlap matrix and $f(E-E_{F})$ is the Fermi distribution 
in the electrodes. Compared to the symmetric SWNT junctions, here the 
Hamiltonian describing the SWNT channel  
$H=H_{0}+\delta V[\delta \rho]+V_{ext}$ includes the contact potential 
$V_{ext}$ (taken as linear voltage ramp here) in addition to the 
charge-transfer induced electrostatic potential change $\delta V$ 
($\delta \rho$ is the density of transferred charge). 

\begin{figure}
\includegraphics[height=3.0in,width=5.0in]{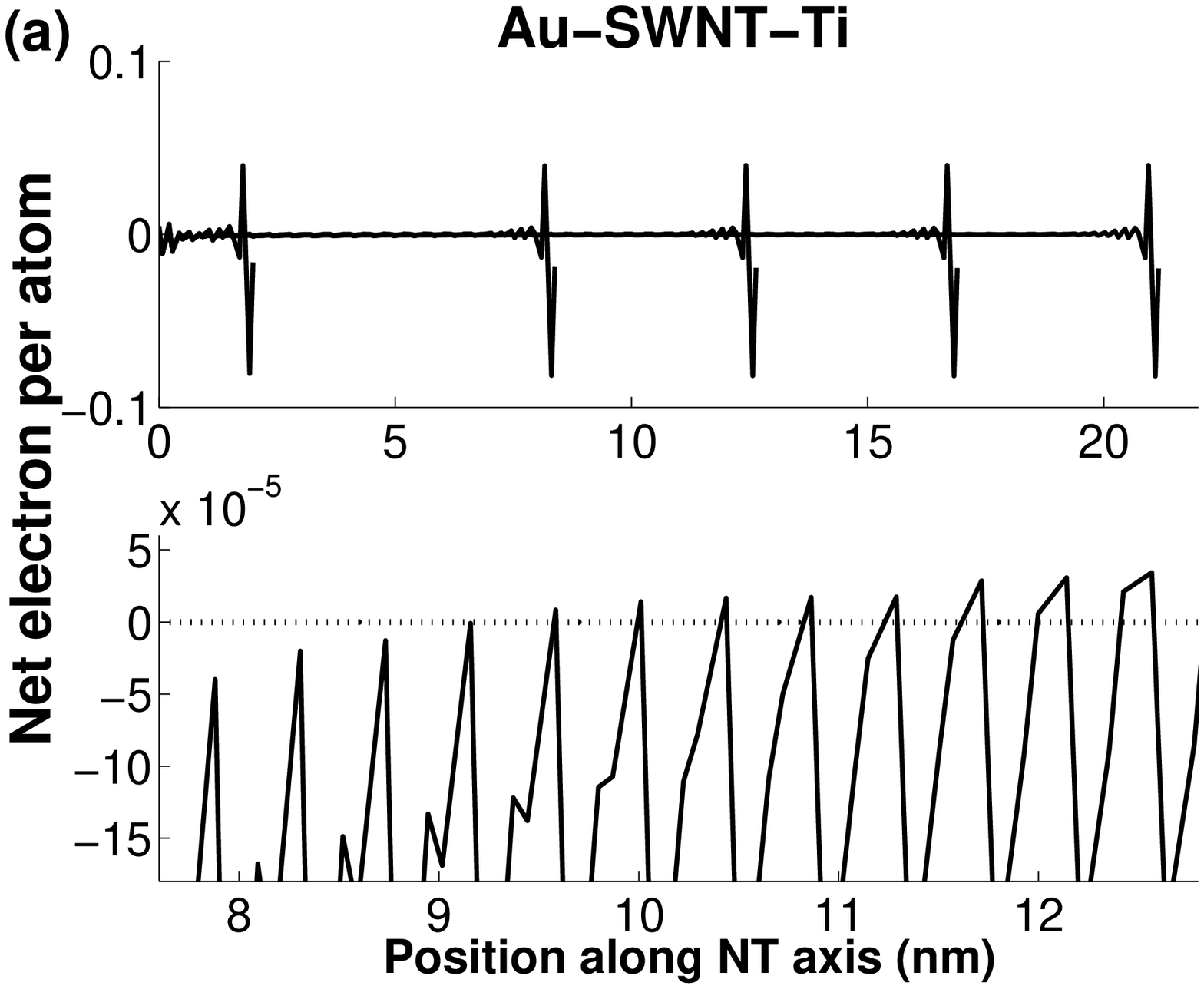}
\includegraphics[height=3.0in,width=5.0in]{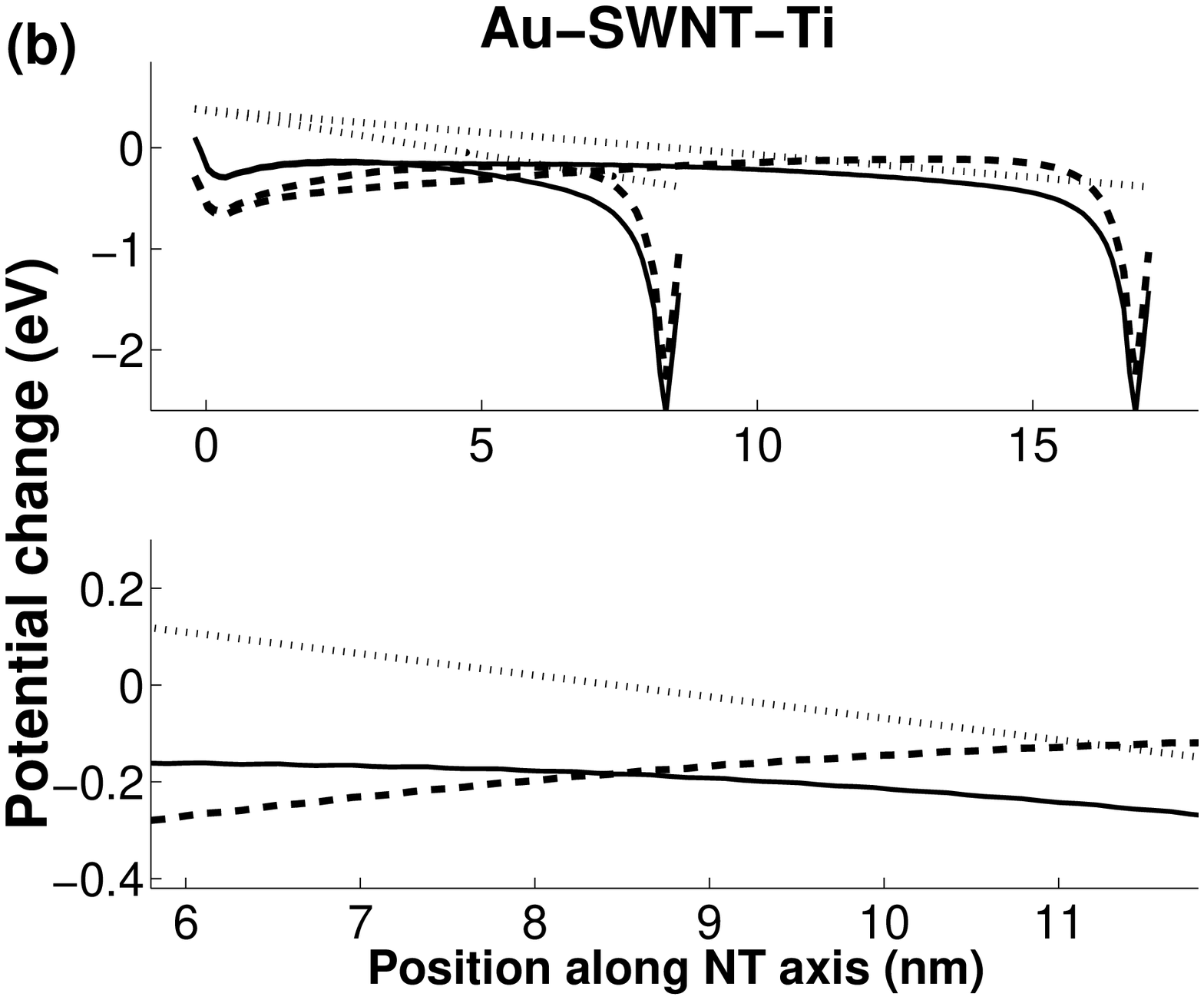}
\caption{\label{xueFig2} Electrostatics of the Au-SWNT-Ti  junction for 
SWNT channels of different lengths. (a) Upper figure shows transferred 
charge per atom as a function of position along the SWNT axis for SWNT 
channel lengths of $2.0, 8.4 12.6, 16.9$ and $21.2$ nm. Lower figure 
shows the magnified view 
of the transferred charge in the middle of the channel for the longest 
(21.2 nm) SWNT studied. (b) Upper figure shows the electrostatic 
potential change at the cylindrical surface of the 20 (8.4 nm) and 40- unitcell 
(16.9 nm) SWNTs studied. The dotted line denote the linear voltage 
ramp $V_{ext}$ (contact potential) due to the work function difference 
of gold and titanium. The dashed line show the 
charge-transfer induced electrostatic potential change 
$\delta V(\delta \rho)$. The solid line shows the net electrostatic 
potential change $V_{ext}+\delta V$. Lower figure shows the 
magnified view of the electrostatic potential change in the 
middle of the 40-unitcell SWNT junction. }
\end{figure}

The calculated charge transfer per atom and electrostatic potential 
change along the cylindrical surface of the SWNT for the Au-SWNT-Ti 
junction are shown in Fig. (\ref{xueFig2}). Previous 
works~\cite{XueNT} have shown that by contacting to 
the high (low)-work function metal Au (Ti), hole (electron) doping is  
induced inside the SWNT channel. Here we find simultaneous electron 
and hole doping inside the SWNT channel for the hetero-metallic 
Au-SWNT-Ti junction (lower figure in Fig. \ref{xueFig2}(a)). 
Since the magnitude of hole doping inside the Au-SWNT-Au junction 
($\approx -5.6 \times 10^{-4}$/atom) is much larger than that of 
the electron doping inside the Ti-SWNT-Ti junction 
($\approx 3 \times 10^{-5}$/atom) due to the larger work function 
difference, the majority of the channel remains hole-doped inside 
the Au-SWNT-Ti junction. Due to the localized nature 
of interface bonding, the charge transfer pattern 
immediately adjacent to the Au(Ti)-SWNT interface remains similar 
to that of the Au-SWNT-Au (Ti-SWNT-Ti) junction both in magnitude 
and shape. The short-wavelength oscillation in the transferred charge 
inside the SWNT channel reflects the atomic-scale variation of charge 
density within the unit cell of the 
SWNT.~\cite{XueNT,Tersoff02} 

\begin{figure}
\includegraphics[height=3.8in,width=5.0in]{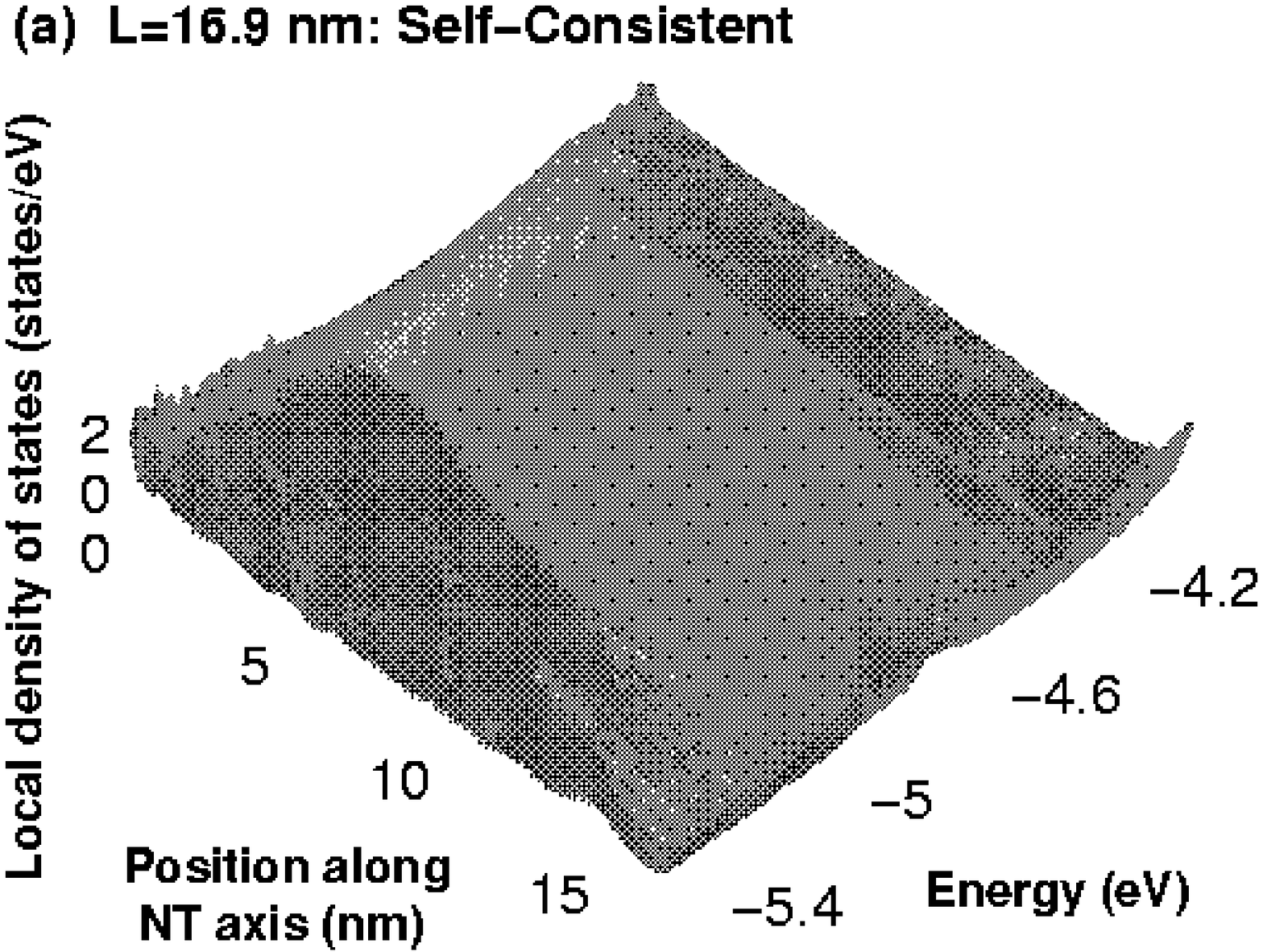}
\includegraphics[height=3.8in,width=5.0in]{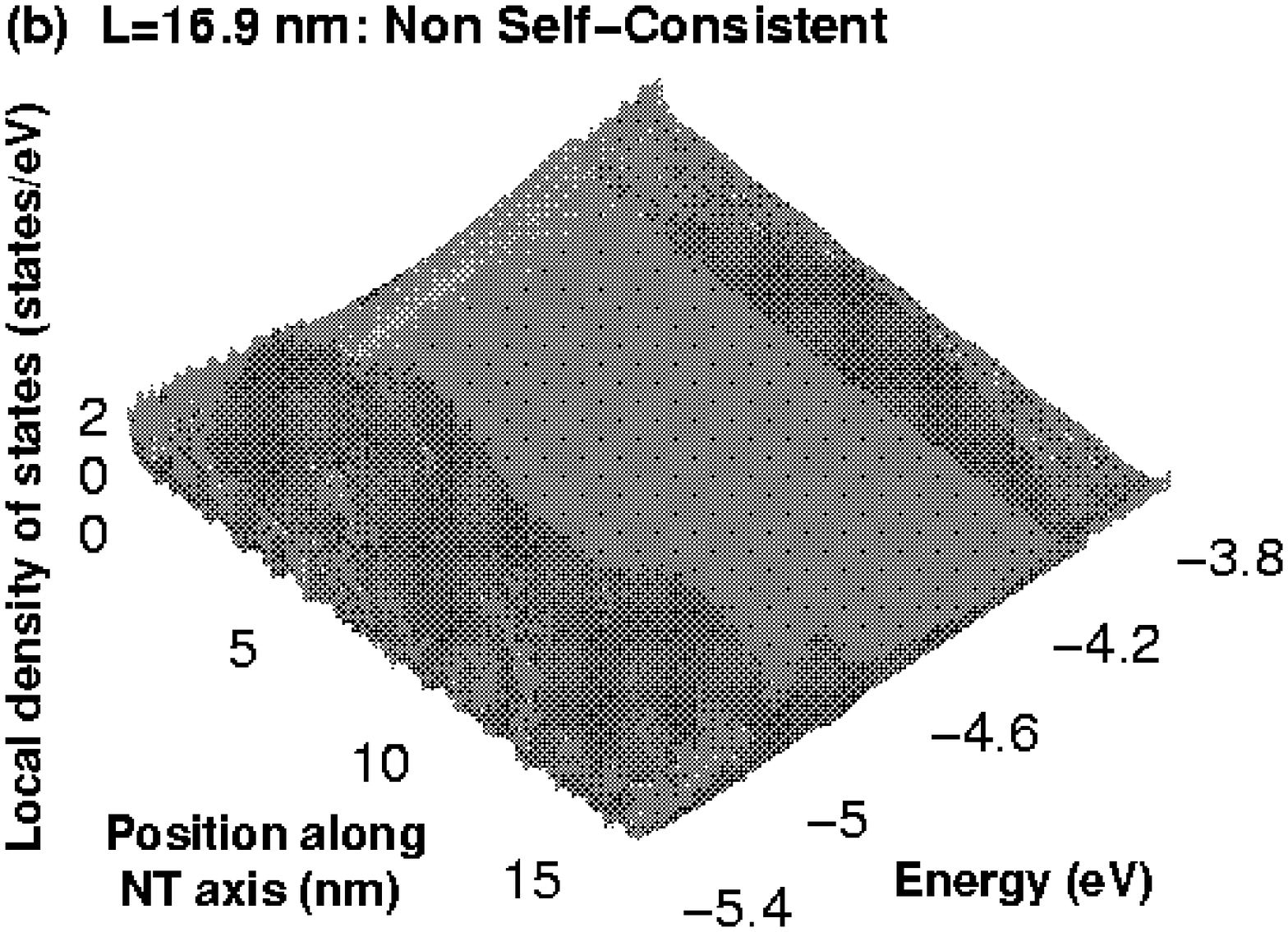}
\caption{\label{xueFig3} (Color online) Local density of states (LDOS) as 
a function of position along the SWNT axis for SWNT channel length 
of $16.9 nm$. We show the result when self-consistent SWNT screening 
of the build-in electric field is included in (a). For comparison we have also 
shown the result for non self-consistent calculation in (b).  The plotted 
LDOS is obtained by summing over the $10$ atoms of each carbon ring 
of the $(10,0)$ SWNT. Note that each cut along the energy axis at 
a given axial position gives the LDOS of the corresponding carbon ring 
and each cut along the position axis at a given energy gives the 
corresponding band shift. }
\end{figure}

\begin{figure}
\includegraphics[height=4.0in,width=5.0in]{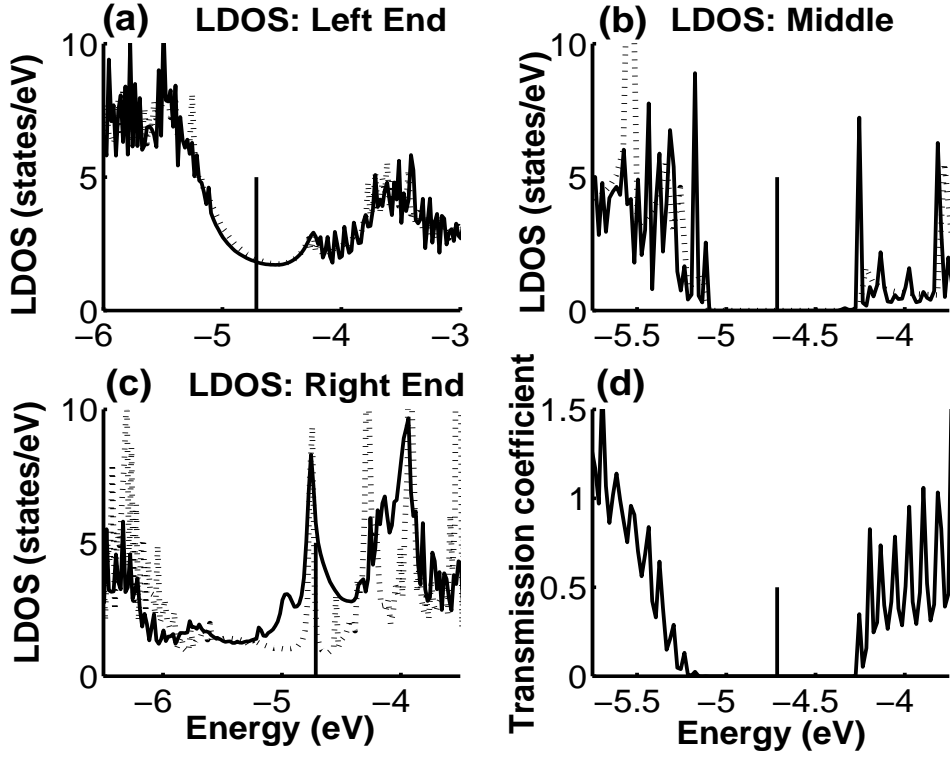}
\caption{\label{xueFig4} Local-density-of-states (LDOS) and transmission 
versus energy (TE) characteristics of the 40-unit cell Au-SWNT-Ti junction. 
(a) LDOS at the 1st unit cell adjacent to the Au side (left end) of 
the Au-SWNT-Ti junction (solid line) 
and the LDOS at the corresponding location of the Au-SWNT-Au 
junction (dashed line). (b) LDOS in the middle unit cell 
of the Au-SWNT-Ti junction (solid line) and the LDOS of the 
bulk (10,0) SWNT (dashed line). (c) LDOS at the 1st unit cell 
adjacent to the Ti side (right end) of the Au-SWNT-Ti junction 
(solid line) and the LDOS at the corresponding location of the 
Ti-SWNT-Ti junction (dashed line). (d) TE characteristics of 
the Au-SWNT-Ti junction. }  
\end{figure}

The contact-induced doping affects the transport characteristics 
by modulating the electrostatic potential profile along the SWNT 
junction. We find that inside the SWNT channel, the built-in electric 
field is screened effectively by the delocalized $\pi$-electron of carbon.  
So the net electrostatic potential change along the cylindrical 
surface ($V_{ext}+\delta V[\delta \rho]$) is much more flat than the 
linear voltage ramp denoting the contact potential except close 
to the metal-SWNT interface (lower figure of Fig. \ref{xueFig2}(b)), 
where its shape remains qualitatively similar to that at the Au (Ti)-SWNT interface 
of the Au-SWNT-Au (Ti-SWNT-Ti) junction. Due to the confined 
cylindrical structure of the SWNT channel, the charge-transfer 
induced electrostatic potential change $\delta V$ decays rapidly 
in the direction perpendicular to the SWNT axis. This has led to a 
different physical picture of band bending in symmetric SWNT 
junctions.~\cite{XueNT} In particular, the band 
lineup inside the SWNT channel has been found to depend 
mainly on the metal work function, while interaction across 
the metal-SWNT interface modulates the band structure close to 
the interface without affecting the band lineup scheme in the 
middle of the channel. Similar physical picture applies to the 
hetero-metallic SWNT junction, 
where we find that the band lineup in the middle of the Au-SWNT-Ti 
junction is essentially identical to that of the SWNT junction with 
symmetric contact to metals with work function of $4.715$ eV.  
This is examined through the local-density-of-states (LDOS) 
of the SWNT channel as a function of position along the SWNT axis in 
Figs. \ref{xueFig3} and \ref{xueFig4}. 

The coupling across the metal-SWNT interface and the corresponding 
strong local field variation immediately adjacent to the Ti-SWNT 
interface has a strong effect on the SWNT band structure there, 
which extends to $\sim 4$ nm away from the interface 
(Fig. \ref{xueFig3}(a)). The band structure modulation at the Au side 
is weaker. For the 40-unit cell (16.9 nm) SWNT, 
the band structure in the middle 
of the SWNT junction remains essentially unaffected. This is shown in 
Fig. \ref{xueFig4}, where we compare the LDOS of the Au-SWNT-Ti 
junction in the left end, the right end and the middle of the SWNT 
channel with the corresponding LDOS of the Au-SWNT-Au, Ti-SWNT-Ti 
junction and the bulk (infinitely long) SWNT wire respectively. Since 
the magnitude of the build-in electric field is smaller than the 
charge-transfer induced local field at the metal-SWNT interface, 
the LDOS at the two ends of the SWNT channel remain qualitatively 
similar to that of the symmetric SWNT junction 
(Figs. \ref{xueFig4}(a) and \ref{xueFig4}(c)). 
Note that the LDOS plotted here has been energetically shifted 
so that the SWNT bands in the middle of the hetero-metallic 
junction line up with those of the symmetric SWNT junctions.   

The above separation of band lineup scheme at the interface and in the 
interior of the SWNT junction implies that in NTFETs, the gate 
segments controlling the device interiors affect the device operation 
through effective modulation of the work function difference between 
the source/drain electrode and the bulk portion of the SWNT channel 
(applying a finite gate voltage to the SWNT bulk leads to an effective 
modulation of its work function relative to the source/drain electrodes), 
while the gate segments at the metal-SWNT interfaces affect the device 
operation by controlling charge injection into the device interior through 
local modulation of the SWNT band structure and Schottky barrier 
shapes including height, width and thickness, in agreement with 
recent lateral scaling analysis of gate-modulation effect~\cite{AvFET3} 
and interface chemical treatment effect in Schottky barrier 
NTFETs.~\cite{MolNT} Note that since the band structure modulation 
at the metal-SWNT interface can extend up to $\sim 4$ nm into the 
interior of the SWNT junction, it may be readily resolved using 
scanning nanoprobe techniques.~\cite{NTSTM}  

The Schottky barrier effect at the metal-SWNT interface can also be analyzed 
through the length-dependent conductance and current-voltage (I-V) 
characteristics of the Au-SWNT-Ti junction, which are calculated 
using the Landauer formula~\cite{XueMol}  
$G=\frac{2e^{2}}{h}\int dE T(E)[-\frac{df}{dE}(E-E_{F})]
=G_{Tu}+G_{Th}$ and $I=\int_{-\infty}^{+\infty}dE \frac{2e}{h}T(E,V)
[f(E-(E_{F}+eV/2))-[f(E-(E_{F}-eV/2))]=I_{Tu}+I_{Th}$ and 
separated into tunneling and thermal-activation contributions 
as $G_{Tu}=\frac{2e^{2}}{h}T(E_{F}),G_{Th}=G-G_{Tu}$ and 
$I_{Tu}=\frac{2e}{h}\int_{E_{F}-eV/2}^{E_{F}+eV/2} T(E,V)dE, 
I_{Th}=I-I_{Tu}$.~\cite{XueNT,XueMol03}  

\begin{figure}
\includegraphics[height=4.0in,width=5.0in]{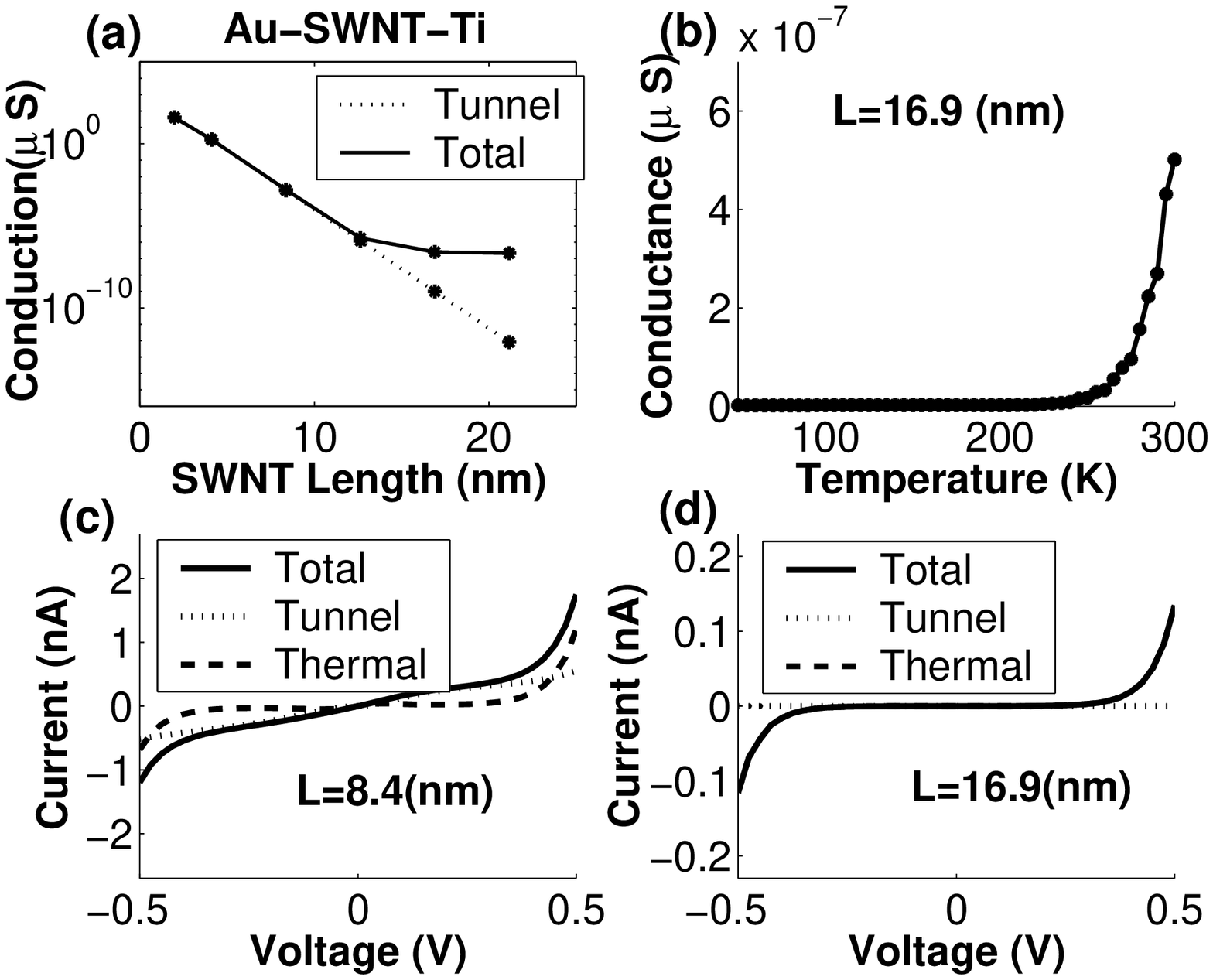}
\caption{\label{xueFig5} (a) Room temperature conductance of 
the Au-SWNT-Ti junction as a function of SWNT channel length. 
(b) temperature dependence of the conductance of the 40-unit cell 
(16.9 nm) SWNT junction. The room temperature current-voltage 
characteristics of the 20- and 40- unit cell SWNT junctions are shown 
in (c) and (d) respectively. }  
\end{figure}

In general the transmission function is voltage-dependent due to the 
self-consistent screening of the source-drain field by the SWNT channel 
at each bias voltage. Since in the case of voltage dropping mostly 
across the interface, the transmission coefficient is approximately 
voltage-independent at low-bias,~\cite{XueMol03} here we calculate 
the I-V characteristics using the equilibrium transmission coefficient 
instead of the full self-consistent calculation at each bias voltage. 
We find that the conductance of the Au-SWNT-Ti junction shows 
a transition from tunneling-dominated to thermal activation-dominated 
regime with increasing channel length, but the length where this occurs is 
longer than those of the symmetric Au/Ti-SWNT-Au/Ti junctions 
(Fig. \ref{xueFig5}(a)). This is partly due to the fact that the 
Fermi-level is closer to the mid-gap of the SWNT band inside the 
channel, partly due to the 
reduced transmission close to the valence-band edge 
(Fig. \ref{xueFig4}(d)) caused by the band structure modulation 
at the Ti-SWNT interface. Due to the finite number of conduction 
channels, the increase of the conductance with temperature is rather 
slow (Fig. \ref{xueFig5}(b)).~\cite{XueNT} The relative contribution 
of tunneling and thermal-activation to the room-temperature I-V 
characteristics is shown in Figs. \ref{xueFig5}(c) and \ref{xueFig5}(d) 
for the 20- and 40-unit cell 
long (8.4 and 16.9 nm) SWNT respectively, where we see that 
thermal-activation contribution increases rapidly with bias voltage 
for the 20-unit cell SWNT junction while the thermal-activation 
contribution dominates the I-V characteristics at all bias voltages for the 
40-unit cell SWNT. 

In conclusion, we have presented an atomistic real-space analysis of 
Schottky barrier effect in the two-terminal SWNT junction with 
hetero-metallic contacts, which shows an effective decoupling of 
interface and bulk effects. Further analysis is needed that treat both 
the gate and source/drain fields self-consistently in the 
real space to achieve a thorough understanding of NTFETs.   


%

\vspace{3.0cm} 

\begin{references}
\bibitem[*]{Xue} Author to whom correspondence should be addressed. 
E-mail: yxue@uamail.albany.edu. URL: http://www.albany.edu/~yx152122.
\bibitem{Dekker} Dekker C 1999 \emph{Phys.\ Today} {\bf 52}(5) 22 
\bibitem{DeMc} Bachtold A, Haley P, Nakanishi T and Dekker C 2001
\emph{Science} {\bf 294} 1317
\bibitem{AvFET} Avouris Ph, Appenzellaer J, Martel R and 
Wind S J 2003 \emph{Proc. IEEE} {\bf 91} 1772 
\bibitem{DaiFET} Javey A, Guo J, Wang Q, Lundstrom M and 
Dai H 2003 \emph{Nature} {\bf 424} 654; Javey A, Guo J, Paulsson M, 
Wang Q, Mann D, Lundstrom M and Dai H 2004 \emph{Phys. Rev. Lett.} {\bf 92} 
106804 
\bibitem{McFET} Yaish Y, Park J-Y, Rosenblatt S, Sazonova V, Brink M 
and McEuen P L 2004 \emph{Phys. Rev. Lett.} {\bf 92} 46401
\bibitem{Barrier} Xue Y and Datta S 1999 \emph{Phys.\ Rev.\ Lett. } 
{\bf 83} 4844; Le{\'o}nard F and Tersoff J 2000 \emph{Phys. Rev. Lett.} 
{\bf 84}  4693; Odintsov A A 2000 \emph{Phys. Rev. Lett. } {85} 150; 
Nakanishi T, Bachtold A and Dekker C 2002 \emph{Phys.\ Rev.\ B} {\bf 66} 73307  
\bibitem{AvFET1} Heinze S, Tersoff J, Martel R, Derycke V, Appenzeller J 
and Avouris Ph 2002  \emph{Phys.\ Rev.\ Lett. } {\bf 89} 106801; 
Appenzeller J, Knoch J, Radosavljevi{\'c} M and Avouris Ph 2004 
\emph{ibid.} {\bf 92} 226802  
\bibitem{XueNT} Xue Y and Ratner M A 2003 \emph{Appl.\ Phys.\ Lett. } 
{\bf 83} 2429; Xue Y and Ratner M A 2004 \emph{Phys. Rev. B} 
{\bf 69} 161402(R); Xue Y and Ratner M A 2004 \emph{Phys. Rev. B} 
In Press (\emph{Preprint} cond-mat/0405465) 
\bibitem{AvFET2} Appenzeller J, Radosavljevi{\'c} M, Knoch J and Avouris Ph 
2004 \emph{Phys. Rev. Lett.} {\bf 92} 48301
\bibitem{NTMe} Minot E D, Yaish Y, Sazonova V, Park J-Y, Brink M and 
McEuen P L 2003 \emph{Phys. Rev. Lett.} {\bf 90} 156401; 
Cao J, Wang Q and Dai H 2003 \emph{Phys.Rev. Lett.} {\bf 90} 157601 
\bibitem{NTCh} Chiu P-W, Kaempgen M, and Roth S 2004 
\emph{Phys. Rev. Lett.} {\bf 92} 246802; Chen G, Bandow S, 
Margine E R, Nisoli C, Kolmogorov A N, 
Crespi V H, Gupta R, Sumanasekera G U, Iijima S and Eklund P C 2003  
\emph{Phys. Rev. Lett.} {\bf 90} 257403  
\bibitem{Note1} Experimentally low-resistance contacts can be 
obtained either by growing SWNT's directly out of the predefined catalyst 
islands and subsequently covering the catalyst islands with metallic 
contact pads (Ref. \onlinecite{DaiFET}) or by using standard 
lithography and lift-off techniques with subsequent annealing at 
high-temperature (Ref. \onlinecite{AvFET1}). In both cases, the 
ends of the long SWNT wires are surrounded entirely by the metals 
with strong metal-SWNT surface chemical bonding, although the exact 
atomic structure of the metal-SWNT interface remains unclear. Contacts 
can also be formed by depositing SWNT on top of the predefined 
metallic electrodes and side-contacted to the surfaces of the metals 
(side-contact scheme), which corresponds to the weak coupling limit 
due to the weak van der Waals bonding in the side-contact geometry 
leading to high contact resistance (Ref. \onlinecite{AvFET}). 
Other types of contact may also exist corresponding to intermediate 
coupling strength. The contact geometry chosen in this work thus 
serves as a simplified model of the low-resistance contact.  
A comprehensive study of the contact effects in SWNT junction 
devices is currently under way and will be reported in a future publication.  
\bibitem{CRC} 1994 \emph{CRC Handbook of Chemistry and Physics} 
(CRC Press, Boca Raton) 
\bibitem{Phonon} Yao Z, Kane C L and Dekker C 2000 \emph{Phys. Rev. Lett.}  
{\bf 84} 2941; Park J-Y, Rosenblatt S, Yaish Y, Sazonova V, Ustunel H, 
Braig S, Arias T A, Brouwer P and McEuen P L 2004  
\emph{Nano Lett.} {\bf 4} 517 
\bibitem{XueMol} Xue Y, Datta S and Ratner M A 2002 \emph{Chem.\ Phys.} 
{\bf 281} 151; See also Datta S 1995 \emph{Electron Transport 
in Mesoscopic Systems} (Cambridge University Press, Cambridge) 
\bibitem{Note2} The SWNT-metal interaction arises from one discrete 
cylindrical shell of metal atoms, surrounded by the bulk metal and treated 
using the Green's function method as detailed in Ref.\ \onlinecite{XueMol}. 
We use a SWNT-metal surface distance of $2.0(\AA)$, close to the average 
inter-atomic spacing in the SWNTs and metals.  
\bibitem{Hoffmann88} Hoffmann R 1988 \emph{Rev.\ Mod.\ Phys.} {\bf 60} 601; 
Rochefort A, Salahub D R and Avouris Ph 1999 \emph{J.\ Phys.\ Chem.\ B} 
{\bf 103} 641   
\bibitem{Tersoff02} Le{\'o}nard F and Tersoff J 2002  \emph{Appl. Phys. Lett.} 
{\bf 81} 4835
\bibitem{AvFET3} Wind S J, Appenzeller J, and Avouris Ph 2003 
\emph{Phys. Rev. Lett.} {\bf 91} 58301
\bibitem{MolNT} Auvray S, Borghetti J, Goffman M F, Filoramo A, 
Derycke V, Bourgoin J P and Jost O 2004 \emph{Appl. Phys. Lett.} 
{\bf 84} 5106 
\bibitem{NTSTM} Freitag M, Radosavljevi{\'c} M, Clauss W and Johnson A T 
2000 \emph{Phys. Rev. B} {\bf 62} R2307; Venema L C, Janssen J W, 
Buitelaar M R, Wild{\''o}er J W G, Lemay S G, Kouwenhoven L P and 
Dekker C 2000 \emph{Phys. Rev. B} {\bf 62} 5238   
\bibitem{XueMol03} Xue Y and Ratner M A 2003 \emph{Phys.\ Rev.\ B} 
{\bf 68} 115406; Xue Y and Ratner M A 2004 \emph{Phys.\ Rev.\ B} 
{\bf 69} 85403  
\end{references}
\end{document}